# Quantitative Ultrasound Perfusion Biomarker: Cross-Modality Validation with Laser Speckle Imaging

Haowei Tai, PhD[1], Oscar Duran-Garnica, DDS, MS[2], Karen Villarreal, DDS, MSD[2], Hsun-Liang Chan, DDS, MS[2], Oliver D. Kripfgans, PhD[1, 3]

[1] Department of Radiology, University of Michigan School of Medicine, Ann Arbor, MI, USA

[2] Division of Periodontology, College of Dentistry, The Ohio State University, Columbus, OH, USA

[3] Department of Biomedical Engineering, College of Engineering, University of Michigan, Ann Arbor, MI, USA

**Corresponding author:** Oliver D. Kripfgans, University of Michigan School of Medicine, 1301 Catherine St., Medical Sciences Bldg 1, Room 6410, Ann Arbor, MI 48109-2026;
Email: greentom@umich.edu

**Abstract**

**Rationale and Objectives:** Tooth extraction is a common dental procedure that often leads to alveolar bone resorption and variable healing outcomes, even when socket preservation techniques involving the placement of biomaterials are applied. Early identification of compromised socket healing remains challenging because current assessment of local tissue perfusion relies largely on subjective visual evaluation. Given the central role of vascular dynamics in tissue regeneration, quantitative and noninvasive assessment of tissue perfusion may provide an objective biomarker for monitoring healing progression. The objective of this study was to develop and validate a quantitative ultrasound processing technique capable of assessing tissue perfusion while minimizing the influence of operator-dependent imaging settings and motion artifacts.

**Materials and Methods:** Color flow ultrasound data were acquired longitudinally from patients (n = 30) undergoing tooth extraction and grafting. Quantitative perfusion parameters, including velocity- and power-weighted color pixel density ($CPD_{vel}$ and $CPD_{pwr}$), were derived from complex-valued color flow (TRI) data using the proposed processing framework within a superficial depth range of 0–1.5 mm from the epithelial surface. The approach was validated using a Quantitative Imaging Biomarker Alliance (QIBA) flow phantom (HE Gel, Gammex, Middleton, WI) to evaluate color gain robustness. Laser speckle contrast imaging (LASCA; PeriCam PSI system, Perimed Inc., Sweden) was performed concurrently for cross-modality comparison.

**Results:** Phantom experiments demonstrated that the proposed method maintained stable velocity quantification across color gain settings ranging from 20 to 100 (a.u) for each tested flow rate (2, 4, 6, and 8 mL/s). *In vivo* analysis demonstrated progressive decreases in perfusion metrics over approximately 2 to 30 weeks following extraction. Nonlinear regression analysis demonstrated a strong longitudinal association for $CPD_{vel}$ ($R^2$ = 0.74) and a moderate association for $CPD_{pwr}$ ($R^2$ = 0.45), while LASCA exhibited a comparable temporal trend ($R^2$ = 0.66). Exponential model fitting further quantified the decay dynamics, yielding half-decay times of 60.3 days for $CPD_{vel}$, 92.4 days for $CPD_{pwr}$, and 48.3 days for LASCA. These values were defined relative to the first post-extraction measurement, which served as the reference time point for decay estimation. In addition, significant linear correlations were observed between LASCA and ultrasound-derived velocity ($R^2$ = 0.68, $p < 0.05$) as well as power ($R^2$ = 0.74, $p < 0.05$).

**Conclusions:** These findings demonstrate the feasibility of a user gain-agnostic quantitative color flow ultrasound framework for noninvasive assessment of tissue perfusion. By extracting perfusion metrics directly from raw TRI data, the method reduces operator dependence and enhances reproducibility across imaging sessions. Quantitative characterization of vascular dynamics within a defined tissue depth range provides objective insight into healing processes not readily captured by conventional qualitative evaluation nor by LASCA. With further validation, this approach holds promise for enabling earlier detection of compromised healing and supporting precision medicine–driven, data-informed clinical decision-making in dental practice.

## Introduction

Tooth extraction is one of the most frequently performed procedures in dentistry, with millions of cases carried out annually worldwide[1,2]. Despite being routine, extraction initiates a cascade of biological events that often result in substantial alveolar bone resorption[3]. Clinical studies have shown that up to 50% of the original alveolar ridge volume may be lost within the first year following extraction[4], potentially compromising future implant placement and prosthetic rehabilitation. To mitigate this bone loss, socket preservation techniques are widely used to maintain ridge dimensions and support subsequent treatment, such as implant placement. However, healing outcomes after socket preservation remain highly variable. A considerable proportion of patients experience compromised graft healing, including persistent inflammation, soft-tissue invagination, delayed integration, or incomplete conversion of graft material into mature bone. These complications may require secondary grafting, prolong treatment, increase costs, and negatively affect patients' quality of life. Therefore, reliable methods are needed to assess healing progression and graft integration at an early stage to support clinical decision-making.

Socket healing after tooth extraction, especially when socket preservation grafting is performed, is a complex multistage process involving coordinated remodeling of soft and hard tissues. It begins with immediate blood clot formation and progresses through inflammation[5], angiogenesis[6], new bone formation[7], and eventual bone maturation and remodeling[8] (**Figure 1**). This process occurs across distinct healing phases including an initial inflammatory stage (approximately weeks 0–2)[9], followed by a prolonged remodeling stage from approximately week 3 to 1 year[10], during which tissue perfusion exhibits phase-dependent changes. Throughout these stages, adequate microcirculation plays a central role. Tissue perfusion and oxygen delivery regulate inflammatory responses, support cellular migration, and facilitate angiogenesis and osteogenesis, making perfusion an important physiological indicator of healing progression. Alterations in microcirculation may occur before clinically visible signs of impaired healing appear, suggesting that vascular assessment could enable earlier detection of complications and more timely intervention. Sustained vascular supply is also essential for maintaining soft-tissue integrity and wound stability and for supporting subsequent bone formation, which remains highly dependent on adequate blood flow. Consistent with this biological role, angiogenesis has been positively correlated with bone volumetric mass, further emphasizing the importance of vascular dynamics in tissue regeneration[11]. Despite its significance, perfusion assessment in periodontal practice largely relies on visual inspection[12] and clinical experience[13]. Such evaluations are subjective and susceptible to inter- and intra-observer variability. An objective imaging modality capable of noninvasively quantifying perfusion would enhance reproducibility, improve diagnostic accuracy, and support more informed clinical decision-making and personalized treatment strategies. Given that disruptions in microcirculation are widely associated with delayed or impaired healing across medical disciplines, local perfusion represents a promising biomarker for monitoring wound recovery in a noninvasive and quantitative manner.

Over the past several decades, multiple imaging techniques have been developed to assess tissue perfusion, including both optical- and ultrasound-based approaches. Optical modalities such as intravital microscopy[14], video capillaroscopy[15], orthogonal polarization spectral imaging[16], and sidestream dark-field imaging provide detailed visualization of capillary morphology and microvascular dynamics. More recently, laser speckle contrast imaging[17] (LASCA) has emerged as a noncontact method for estimating superficial blood flow and generating high-resolution surface perfusion maps. Despite these advances, the clinical application of optical techniques in periodontal and extraction socket healing remains limited.

Most optical methods have a shallow penetration depth and restricted access to posterior intraoral regions. Some approaches are invasive or highly sensitive to motion artifacts, and device size and positioning constraints can further limit practical use within the oral cavity. Importantly, vascular remodeling within gingival and periodontal connective tissues plays a critical role in soft-tissue repair and restoration of functional architecture following injury[18,19]. Because these processes extend beyond the superficial capillary plexus, perfusion dynamics at greater tissue depths may provide additional information about healing progression that is not captured by surface-based imaging alone. These limitations highlight the need for complementary imaging modalities capable of providing reliable perfusion measurements across varying tissue depths and anatomical locations. Access to deeper tissue perfusion may offer additional insight into healing mechanisms and help guide the development of more targeted therapies.

Ultrasound imaging offers several practical advantages for dental applications, including real-time acquisition, noninvasive operation, and the ability to probe subsurface tissues beyond the penetration limits of optical methods. Recent advances in transducer miniaturization have enabled the development of probes specifically designed for intraoral use, with apertures as small as approximately 16.2 × 3 mm, making ultrasound imaging feasible within the confined anatomical space of the oral cavity[20,21]. This hardware innovation[22,23] has opened new possibilities for applying ultrasound to dental and periodontal imaging.

Ultrasound techniques can provide indirect estimates of tissue perfusion, including flow velocity and relative perfusion volume, or power, by analyzing phase shifts in backscattered signals from moving red blood cells. However, reproducible quantitative assessment in clinical dental imaging remains challenging because color flow–based velocity and power measurements are sensitive to acquisition settings. Parameters such as color gain are routinely adjusted to optimize image quality for each patient. These adjustments can affect the quantitative accuracy of perfusion measurements because they alter the appearance and spatial extent of the displayed color flow. Without a processing approach that reduces sensitivity to these system settings, this variability limits reproducibility and interferes with reliable inter- and intra-operator comparisons, thereby constraining the clinical utility of ultrasound as a quantitative perfusion assessment tool.

To address these limitations, the present study proposes an ultrasound signal-processing technique designed to quantify local perfusion within a superficial depth range of 0–1.5 mm while minimizing the influence of system-dependent parameters. We hypothesize that this processing approach can reliably capture perfusion-related information within this depth range and enable quantitative assessment of perfusion changes over time following tooth extraction. If validated, this method may help transition color flow ultrasound from a predominantly qualitative imaging modality to a quantitative tool for monitoring healing dynamics, with potential implications for improving clinical evaluation of extraction socket healing and bone maturation.

## Materials and Methods

### Ultrasound Data Acquisition

The multi-site study was approved by the University of Michigan Medical School as the IRB of record (HUM00226516). All subjects had to give written informed consent as the study followed the ethical guidelines of the Helsinki Declaration. Ultrasound complex-valued color flow (TRI) data were acquired from patients (n = 30) at The Ohio State University (OSU) College of Dentistry, Division of Periodontology, during longitudinal follow-up visits using a clinical ultrasound system (ZS3, Mindray, USA) equipped with a small-footprint, 128-element linear array probe (L30-8, Mindray, USA). The transducer has a center frequency of 18 MHz and was operated in CSH24 mode, i.e., with spatial compounding and harmonic imaging (12–24 MHz) enabled. During ultrasound data acquisition, the clinician was permitted to adjust scanner controls, including gain, pulse repetition frequency (PRF), imaging plane, and depth, to obtain optimal display images. A virtual convex configuration was applied to the linear transducer, with scanlines distributed across ±20°, and angle compounding was maintained at the default setting. Longitudinal ultrasound imaging was performed to monitor healing of the extraction socket over time. Beamformed and post auto-correlation TRI cine-loop data were acquired from the facial aspect of the edentulous ridge. Each patient had seven scheduled visits: V1, pre-procedure assessment; V2, tooth extraction and graft placement, day 0; and V3–V7, recovery and healing monitoring, 2–30 weeks. The ultrasound processing pipeline was further validated using a Quantitative Imaging Biomarker Alliance (QIBA) flow phantom (HE Gel, Gammex, Middleton, WI). The phantom was customized to committee specifications (28 cm × 30.5 cm × 22 cm; sound speed, 1540 m/s; attenuation, 0.5 dB/MHz/cm) and allowed control of flow rates[24]. This phantom study evaluated the effects of different color flow metric gain settings (20, 40, 60, 80, and 100) on velocity and power measurements and assessed both inter- and intra-gain variability. For each gain and flow condition, measurements were repeated five times to evaluate measurement consistency. Flow rates were set at 2, 4, 6, and 8 mL/s.

**Ultrasound Data Preprocessing**

All ultrasound data underwent scan conversion and motion tracking to recover the physical geometry of the imaged tissue and its motion during acquisition. For each cine loop, an expert manually delineated the region of interest (ROI) encompassing the free gingiva and epithelium on a representative frame **(Figure 2)**. Offline motion correction based on corner detection and optical flow was applied to compensate for tissue displacement during the 6-second image acquisition and to propagate the ROI accurately across all frames. Pixel-level velocity was estimated from the phase of the exported complex-valued color flow (TRI) signal. The phase was converted to physical velocity using the scanner-displayed velocity range. Power was derived from the TRI magnitude and scaled based on the scanner-reported power mapping. To reduce noise, velocity outliers were excluded in regions where the corresponding power was below 17.5 dB (maximum power was approximately 35 dB). The epithelial surface was estimated from the epithelium ROI after ROI propagation. A centerline was computed using skeletonization, and its superficial boundary was defined as the topline. Pixel depth was then calculated as the perpendicular distance from this surface for each pixel within the ROI. Velocity and power values within the ROI were grouped by depth and averaged to obtain mean depth-resolved metrics using the previously described velocity- and power-weighted color pixel density methods ($CPD_{vel}$ and $CPD_{pwr}$)[25]. This enabled evaluation of depth-dependent changes in color flow information. Finally, color area was defined as the percentage of pixels within the ROI exceeding the predefined power threshold of 17.5 dB, normalized to the maximum observed color area across all measurements.

**Optical Imaging**

Laser speckle contrast analysis (LASCA; PeriCam PSI system, Perimed Inc., Sweden) was performed concurrently to assess tissue perfusion. LASCA provides high spatial resolution, approximately 10 μm, and high temporal resolution, up to 113 frames per second, generating two-dimensional color-coded maps of relative perfusion intensity. The technique quantifies superficial tissue blood flow, with a penetration depth up to approximately 2 mm, by analyzing speckle patterns generated by scattered near-infrared laser light, wavelength approximately 785 nm, projected over a defined two-dimensional region of interest. Light reflected from stationary structures produces a stable speckle pattern, whereas motion from flowing blood cells induces speckle fluctuations that decorrelate proportionally to flow velocity. LASCA imaging was performed at each study visit. For normalization, the ratio of mean perfusion at the free gingival extraction site[26–28] to that of adjacent healthy tissue at the same vertical level was calculated to reduce intersubject variability. Imaging was conducted in 30 patients across all study visits, weeks 2–30. All patients had both LASCA and ultrasound data available, except one patient who had only ultrasound data.

**Statistical Analysis**

Data normality was assessed using the Shapiro–Wilk test. Normally distributed data were presented as mean ± standard deviation (SD), whereas non-normally distributed data were expressed as median and interquartile range. To facilitate temporal comparisons and account for unbalanced longitudinal data, ultrasound-derived perfusion metrics and LASCA measurements were grouped into 15-day intervals beginning at the 2-week postoperative visit, with each patient's extraction visit defined as day 0. Linear regression analysis was performed to evaluate the association between ultrasound- and LASCA-derived measurements, and corresponding p-values were used to assess statistical significance. In addition, nonlinear regression analysis was conducted to characterize the temporal evolution of ultrasound-derived perfusion metrics and LASCA perfusion ratios. Nonlinear exponential models were applied to evaluate changes in $CPD_{vel}$ and $CPD_{pwr}$ as a function of time. The coefficient of determination ($R^2$) was used to assess goodness of fit. A p-value less than 0.05 was considered statistically significant. All data visualizations were generated using Prism software (GraphPad Software, San Diego, CA).

## Results

### Phantom Validation of Quantitative Accuracy

The impact of ultrasound gain on velocity measurements is illustrated in **Figure 3** using the QIBA flow phantom at a fixed flow rate of 2 mL/s. As gain increased from 40 to 80, the displayed color region progressively expanded, with saturation becoming evident at a gain of 100. In contrast, velocity quantification using the proposed color flow signal-processing technique remained stable across gain settings, indicating robustness to gain-dependent display variations.

To further evaluate performance, the technique was tested on the QIBA phantom using combinations of gain (20, 40, 60, and 80) and flow rate (2, 4, 6, and 8 mL/s). The results indicate that increasing flow rate was associated with both elevated color-encoded velocity measurements and expansion of the colored flow region when gain was held constant. Conversely, when the flow rate was fixed and the gain was varied, the color maps showed minimal visual differences in appearance or spatial distribution (**Figure 4**). This pattern was consistent across all tested flow rates. The number of color pixels increased with flow rate as the colored region expanded within the ROI, with little further change observed between 6 and 8 mL/s. This behavior was independent of color flow gain.

These trends are more clearly demonstrated in the quantitative analysis shown in **Figure 5**. Solid lines with circular markers represent measured velocity values across different flow rates and gain settings, plotted on the left y-axis, with color indicating gain level. Red dashed lines represent the percentage of color area within the ROI, plotted on the right y-axis. Color flow images showed that the extent of the colored region within the ROI increased with flow rate. As flow increased from 2 to 8 mL/s, the color area within the ROI increased from approximately 34%–36% to 75%–80%. Only minimal differences were observed between 6 and 8 mL/s, approximately 75%–77% versus 75%–80%. For a given flow rate, the percentage of color area showed no significant differences across gain settings.

Velocity measurements exhibited similar behavior across gain settings for each flow condition, indicating gain-invariant quantification. For flow rates of 2, 4, 6, and 8 mL/s, repeated measurements were performed at each gain setting. The resulting mean velocity values ranged from 0.34 to 0.35, 0.97 to 1.02, 1.50 to 1.54, and 1.82 to 1.87 across gain settings of 20, 40, 60, and 80. These ranges suggest slightly lower variability at flow rates of 2 and 8 mL/s than at intermediate flow rates of 4 and 6 mL/s, although overall consistency was maintained across all experimental conditions. No significant differences were observed when imaging the same flow rate under different gain settings ($p > 0.05$).

### *In Vivo* Quantification of Tissue Perfusion

The proposed technique was subsequently applied to *in vivo* ultrasound data acquired from the free gingiva beginning at the 2-week postoperative visit. A progressive reduction in the extent of the color flow region was observed from approximately 2 to 30 weeks after tooth extraction. The final time point represented the latest post-extraction observation. This trend was consistent with both $CPD_{vel}$ and $CPD_{pwr}$ measurements. By approximately 30 weeks after tooth extraction, both parameters approached minimal levels. This likely reflects tissue healing and reduced vascular activity (**Figure 6**). Together, these findings suggest a gradual decrease in blood volume at the extraction site over time.

Representative LASCA images acquired from the same patient across multiple visits are shown in **Figure 7**. ROIs were selected at two locations within the gingival tissue: the extraction site and an adjacent healthy region. Over time, the LASCA color map transitioned from red to light blue at the extraction site, indicating reduced perfusion. This healing progression is also evident in the corresponding clinical photographs, where the implanted membrane is visible at 2 weeks post-extraction and gradually resolves at subsequent time points. The perfusion ratio was calculated using these two ROIs, as indicated by the black boxes.

Finally, all measurements were summarized to evaluate the relationships between LASCA and ultrasound-derived metrics (**Figure 8**). Exponential regression analysis characterized the temporal behavior of each parameter, revealing a strong longitudinal association between $CPD_{vel}$ and time ($R^2 = 0.74$; **Figure 8A**) and a moderate association between $CPD_{pwr}$ and time ($R^2 = 0.45$; **Figure 8B**). Linear regression analysis demonstrated significant correlations between LASCA and $CPD_{vel}$ ($R^2 = 0.68$, $p < 0.05$; **Figure 8C**) as well as $CPD_{pwr}$ ($R^2 = 0.74$, $p < 0.05$; **Figure 8D**). As an independent imaging modality for comparison, LASCA exhibited a comparable nonlinear temporal trend ($R^2 = 0.66$; **Figure 8E**). Exponential model fitting further quantified the decay dynamics of each parameter. Estimated half-decay times were 60.3 days for $CPD_{vel}$, 92.4 days for $CPD_{pwr}$, and 48.3 days for LASCA.

## Discussion

This study demonstrates the feasibility of using quantitative ultrasound to noninvasively assess tissue perfusion during healing following tooth extraction. Because perfusion reflects tissue healing status[29,30], objective measurements could improve treatment planning and clinical decision-making[31,32] while helping reduce unnecessary or premature interventions[33–35]. To address the current lack of reliable quantitative tools, we developed an ultrasound-based signal-processing framework that measures perfusion within a defined superficial tissue depth while reducing sensitivity to scanner-dependent imaging settings. By enabling consistent comparisons across imaging settings and longitudinal clinical visits, this approach may minimize operator-induced variability and expand the clinical utility of ultrasound for quantitative perfusion assessment in dental practice[34].

A practical advantage of ultrasound compared with optical techniques such as LASCA is its ability to probe deeper tissue layers. While LASCA provides high-resolution surface perfusion mapping, it is limited to superficial vascular beds. Ultrasound, in contrast, can assess blood flow within subepithelial and deeper connective tissue structures, potentially revealing healing dynamics not visible with surface-based imaging alone. In addition, the compact intraoral probe used in this study facilitates access to posterior oral regions, where optical devices may be difficult to position.

The robustness of the proposed quantification approach was evaluated using a QIBA flow phantom[36], which allows digital control of flow rate, minimizing confounding factors and other sources of measurement variability. Conventional color flow display showed progressive expansion of the color region as gain increased, with color blooming observed at the highest gain setting, i.e., 100. Such display dependence can aid perfusion visualization but may also confound quantitative interpretation and compromise longitudinal comparisons. In contrast, $CPD_{vel}$ measurements derived from the proposed processing pipeline remained stable across gain settings when the flow rate was held constant. This supports the reliability of quantitative comparisons across imaging sessions and operators. Although variations were observed at intermediate flow rates, measurements remained consistent across gain settings, and no significant differences ($p > 0.05$) were detected among repeated measurement groups at any flow rate. This suggests that the proposed technique effectively mitigates display-related bias. By eliminating gain dependency, the proposed method enhances the feasibility of quantitative and qualitative comparisons across time and between operators.

*In vivo* application of the method further supports its clinical relevance. We focused on the free gingival region and evaluated perfusion within 0–1.5 mm from the epithelial surface. This depth range was selected based on prior literature[37–39], which suggests that the superficial gingival vascular network is confined to within a few millimeters of the tissue surface. The selected range of 0–1.5 mm lies within this region and falls within the reported LASCA penetration depth of up to approximately 2 mm when wavelengths are between 658 and 850 nm[38]. Ultrasound-derived perfusion metrics demonstrated a progressive decrease over the period from week 2 to week 30 following tooth extraction. This temporal pattern was consistent with LASCA measurements and aligns with established wound healing physiology[40,41], in which vascular activity decreases as inflammation resolves and tissue remodeling progresses. By 27–30 weeks, both velocity and power metrics approached minimal levels, indicating markedly reduced vascular activity. This may reflect physiologic vascular regression in mature and healed tissue, while reduced detectability due to perfusion below the sensitivity threshold of the current imaging system cannot be excluded in rare cases[42].

The observed correlation between ultrasound and LASCA measurements further supports the validity of the proposed approach. Linear regression analysis revealed statistically significant associations between LASCA perfusion ratios and both ultrasound $CPD_{vel}$ and $CPD_{pwr}$ metrics, with a stronger correlation observed for power measurements. This may be due, in part, to the fact that power measurements more directly reflect the concentration of moving scatterers and overall blood volume, whereas velocity measurements primarily capture flow speed[43,44]. Although ultrasound and LASCA measure perfusion using different physical principles and sampling geometries, the convergence of findings across modalities strengthens confidence in the reliability of the quantitative ultrasound framework.

Beyond monitoring, quantitative perfusion assessment may have broader clinical implications. Objective evaluation of vascular response could aid in early identification of delayed healing, excessive inflammation, or compromised blood supply. In dental practice, such information may inform decisions regarding implant placement timing, membrane management, or adjunctive therapies. If further validated, this approach could support a shift from qualitative observation toward data-driven assessment, enabling more personalized treatment strategies. The ability to consistently quantify perfusion from raw ultrasound data also opens opportunities for longitudinal monitoring and potentially for predictive modeling of healing trajectories.

Several limitations should be acknowledged. First, although phantom calibration demonstrated robustness to gain variation, the power threshold used to exclude low-signal regions was selected empirically. Optimization of this threshold warrants further investigation in larger and more diverse cohorts. Adaptive thresholding strategies may improve sensitivity while preserving noise suppression, but such approaches require additional validation. Second, ultrasound and LASCA measurements were not geometrically co-registered. LASCA provides two-dimensional surface perfusion maps[45–47], whereas ultrasound samples perfusion within a two-dimensional cross-sectional imaging plane inside the tissue[48]. Differences in sampling geometry and depth sensitivity may partially account for the moderate correlations observed, particularly the stronger association between LASCA and power compared with LASCA and velocity. Future studies incorporating multimodal image registration or volumetric ultrasound acquisition may help clarify spatial relationships between modalities and improve cross-platform comparability. Additionally, this study focused on a localized gingival region within a defined healing timeframe. Expanding the analysis to other intraoral sites, pathological conditions, and longer follow-up periods would provide a more comprehensive evaluation of clinical applicability. Although the present findings demonstrate feasibility and internal consistency, larger cohort studies are needed to establish normative reference ranges and to assess diagnostic performance metrics such as sensitivity and specificity.

Despite these limitations, the present study establishes a quantitative framework for ultrasound-based tissue perfusion assessment that remains stable across variations in clinical imaging conditions. By leveraging TRI data and a structured signal-processing pipeline, the proposed approach reduces operator dependence and improves measurement reproducibility. The consistent findings across phantom validation, *in vivo* temporal trends, and cross-modality comparison with LASCA collectively support the robustness of the methodology and highlight its potential as a reliable tool for monitoring tissue healing.

## Conclusions

This study developed a quantitative ultrasound processing technique based on raw color flow data to minimize the influence of operator-dependent clinical imaging settings. Perfusion within a defined superficial depth range of 0–1.5 mm from the epithelial surface was quantified. The averaged velocity and power measurements showed a statistically significant correlation with the optical imaging modality LASCA ($p < 0.05$). In addition, both velocity and power demonstrated nonlinear longitudinal changes following extraction ($R^2 = 0.74$ and 0.45, respectively), while LASCA exhibited a comparable nonlinear trend ($R^2 = 0.66$). These findings indicate that quantitative ultrasound can detect time-dependent changes in gingival perfusion during healing. The ability to noninvasively assess vascular dynamics across superficial and subsurface tissue layers may provide meaningful insight into the biological processes underlying tissue recovery. With further validation in larger cohorts and continued methodological refinement, this technique has the potential to expand the role of ultrasound in dentistry from a primarily qualitative monitoring tool to a more comprehensive quantitative diagnostic and clinical decision-support modality.

**Acknowledgement**

This research was supported by the National Institutes of Health (NIH) R01DE030872. The authors declare no conflicts of interest.

**Data availability statement**

The data that support the findings of this study are not publicly available due to ethical and privacy restrictions but are available from the corresponding author upon reasonable request.

## Figure Legends

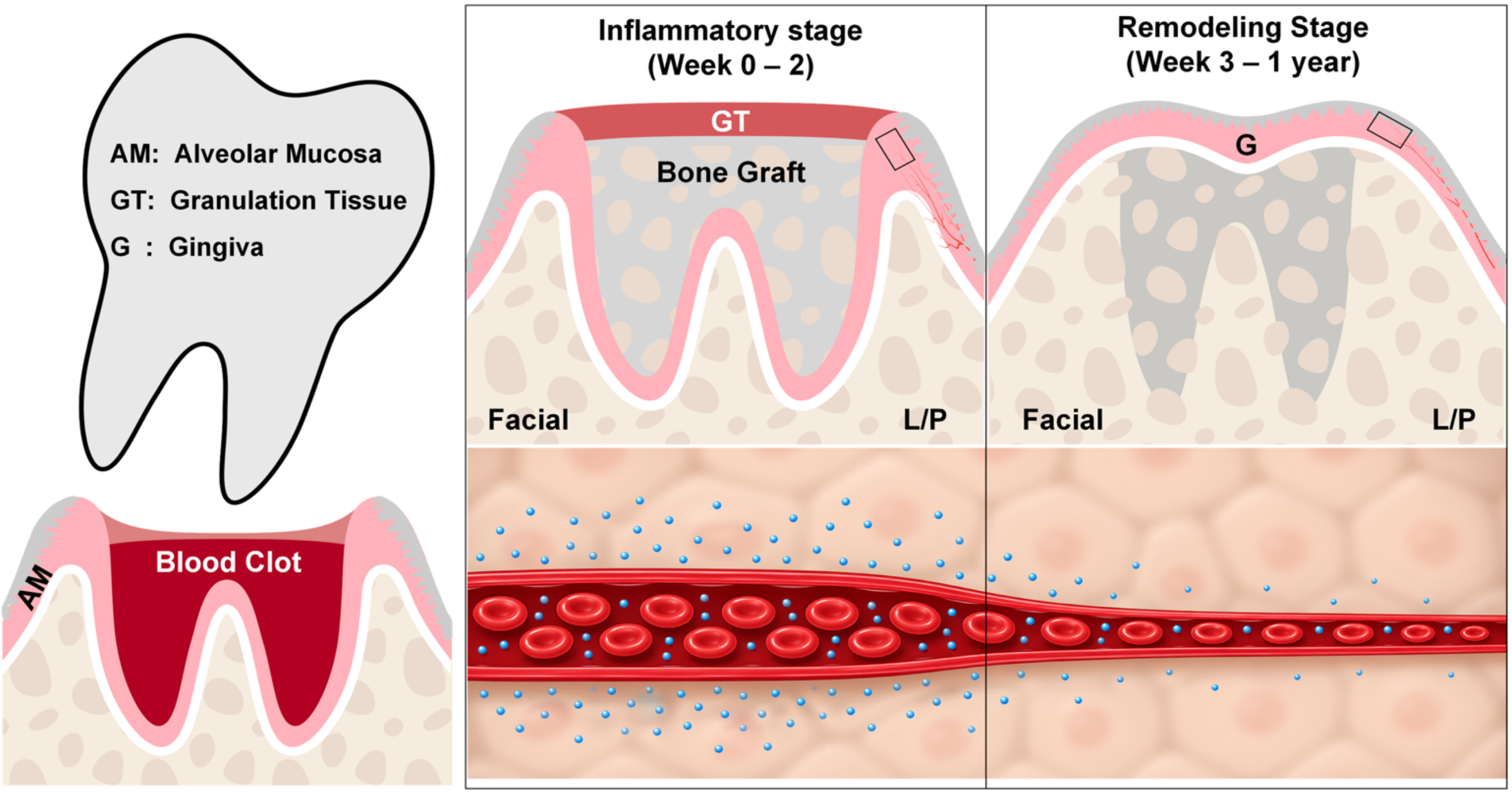

**Figure 1.** Angiogenesis plays a critical role in oral wound healing and is essential for tissue regeneration following tooth extraction. Vascular development is closely linked to bone repair, as adequate blood supply delivers oxygen and immune cells that support tissue integrity and remodeling. Immediately after extraction, vascular disruption occurs, followed by progressive reorganization of the microvasculature as perfusion demands change across different stages of healing. These vascular changes may serve as potential biomarkers of healing progression and could help guide personalized treatment planning while improving understanding of the underlying biological mechanisms. High-frequency ultrasonography is hypothesized to capture these vascular changes and provide a quantitative assessment tool (L/P = lingual in the mandible and palatal in the maxilla).

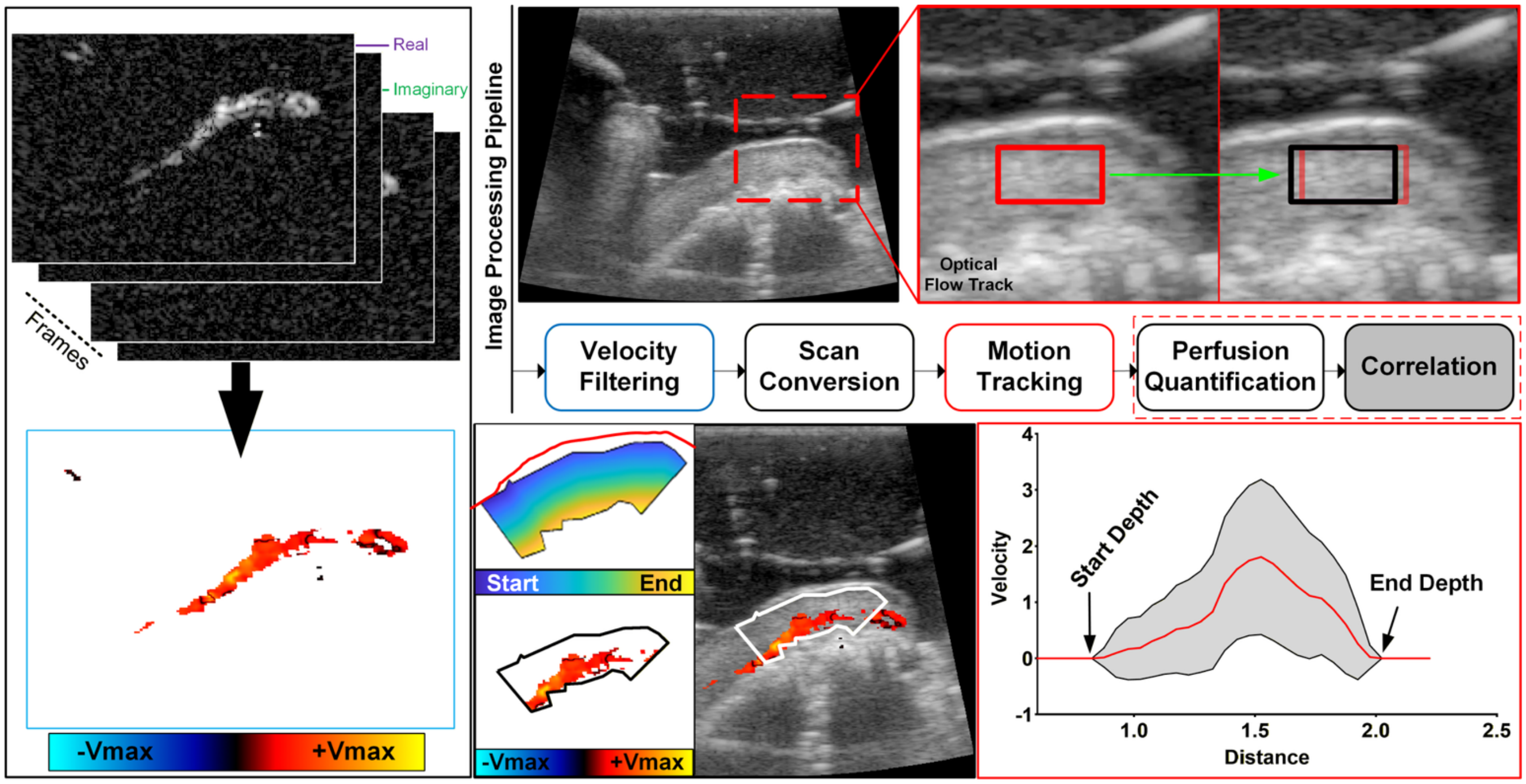

**Figure 2.** Flowchart of the ultrasound signal-processing and velocity- and power-weighted color pixel density (CPD) quantification pipeline. Longitudinal beamformed complex-valued color flow (TRI) data provided phase-shift information, enabling estimation of blood flow velocity based on the system-reported velocity range. Color pixels with power values below 17.5 dB were excluded to reduce noise. Motion tracking compensated for tissue displacement during cine loop acquisition. The free gingival region (white ROI) was manually selected, and the distance of each pixel from the epithelial surface was calculated. Velocity and power values were subsequently analyzed as functions of depth for quantitative assessment, with the mean (red line) and standard deviation (SD; gray region) shown in the bottom-right panel.

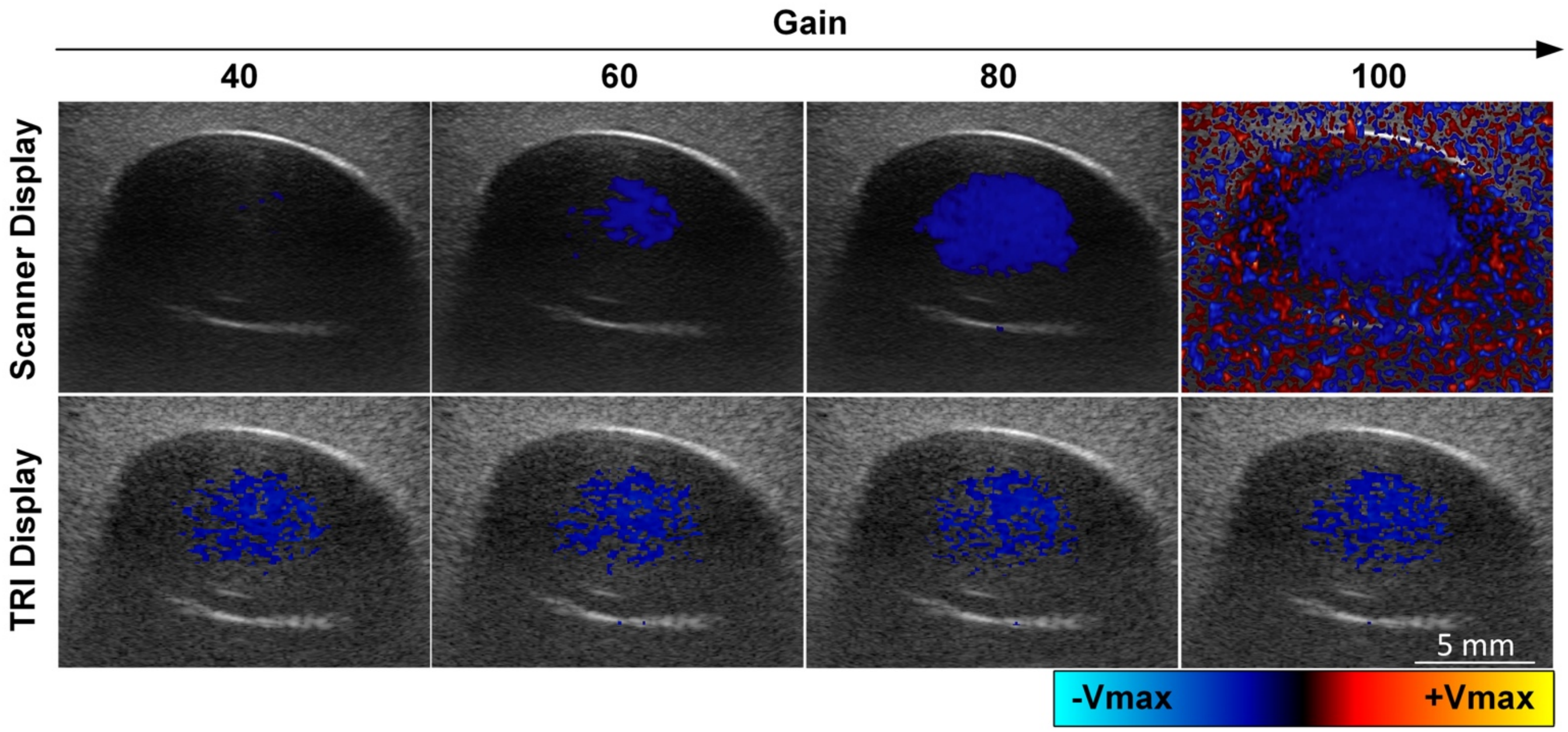

**Figure 3.** Comparison of conventional ultrasound display images (top) and images derived from raw TRI data (bottom). The color flow receive gain setting substantially affects the appearance and spatial extent of the color region in the conventional display image, whereas the complex TRI data–based processing remains largely insensitive to gain variation. As gain increases from 40 to 100 (left to right), the color region in the conventional display progressively expands, while the TRI data–derived quantification remains stable with no apparent change in color area. Data were acquired at a fixed flow rate of 2 mL/s.

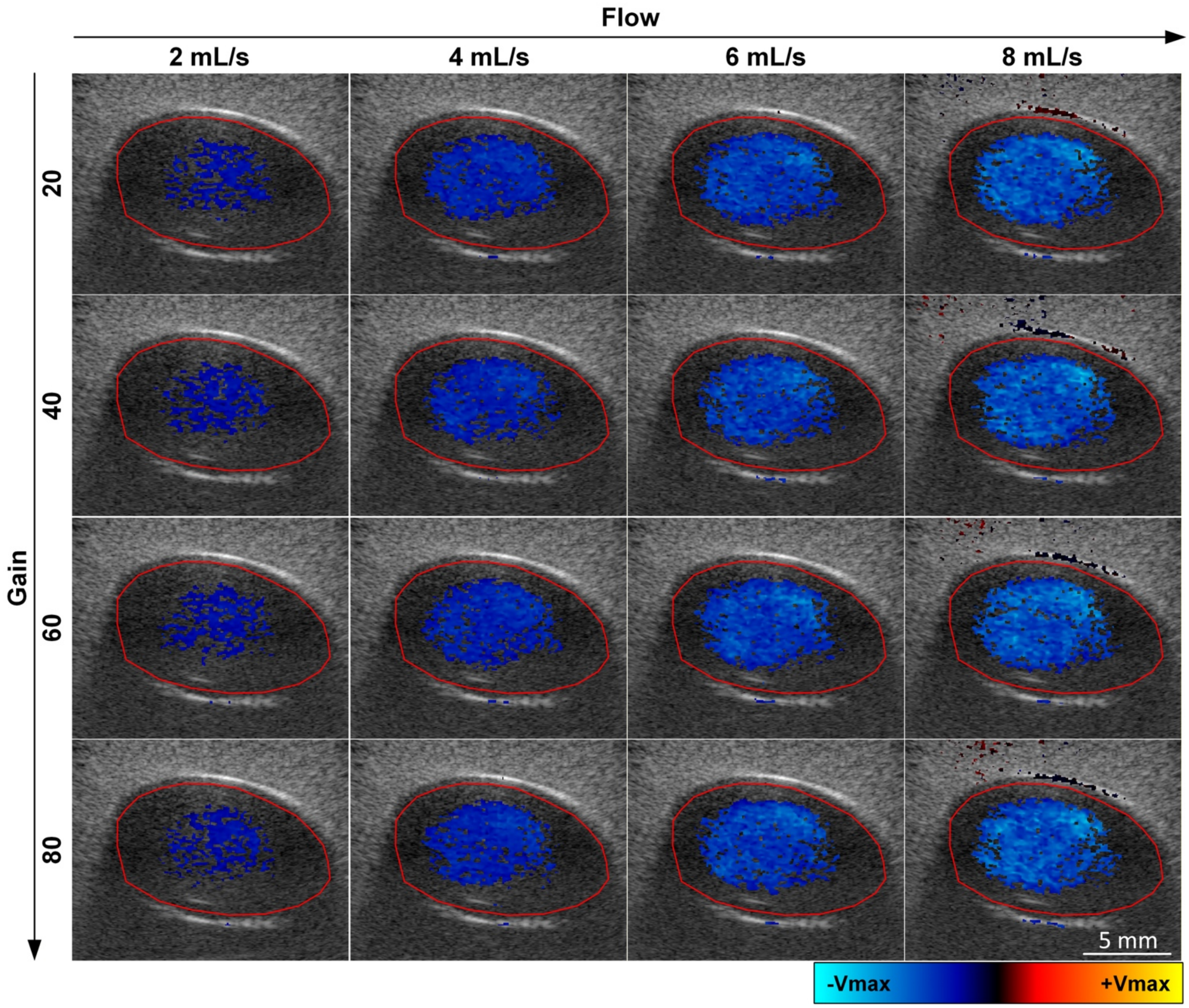

**Figure 4.** Comparison of the effects of gain and flow rate on velocity estimation. Each row represents a fixed gain setting, and each column represents data acquired at a fixed flow rate. When gain is held constant, within a row, both color intensity and spatial extent increase progressively as flow rate rises from 2 to 8 mL/s. This behavior is expected because the color velocity scale is held constant and the wall filter suppresses low-velocity signals. In contrast, when flow rate is held constant, within a column, no visible differences in the color region are observed as gain varies from 20 to 80.

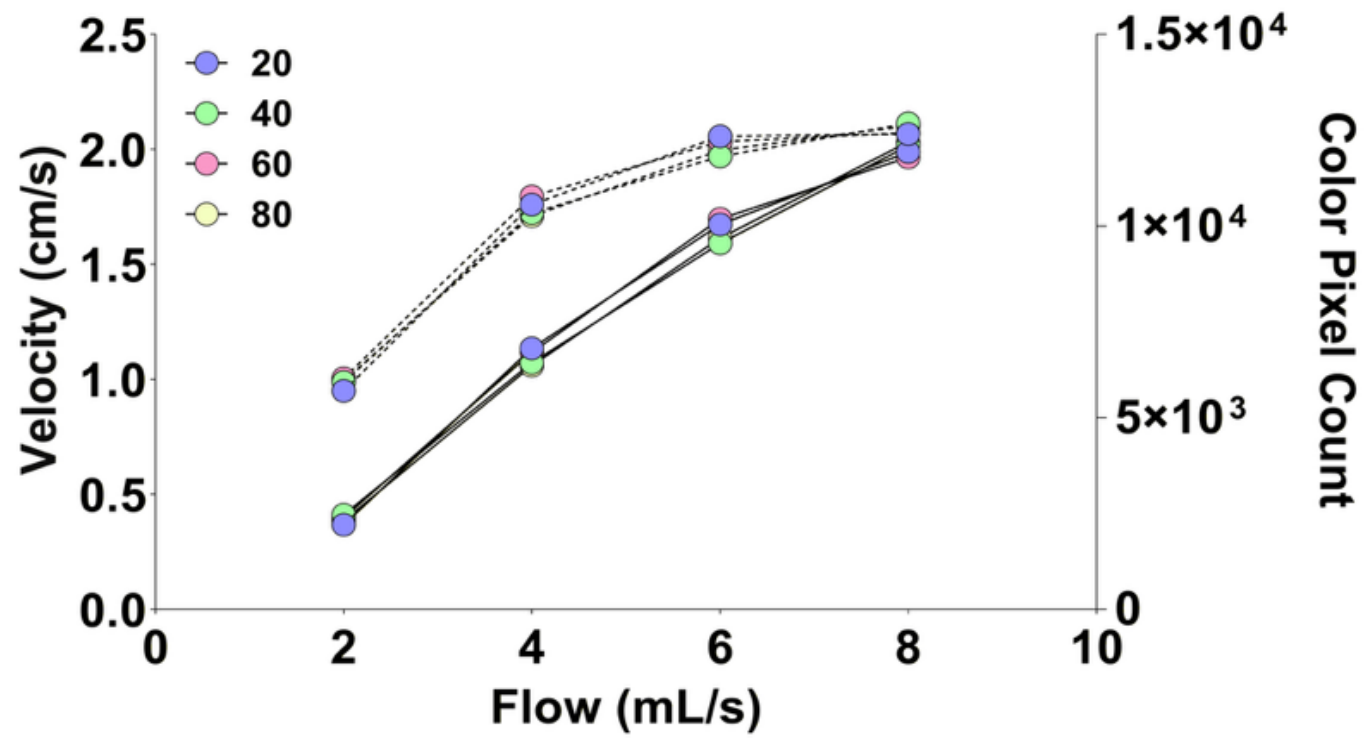

**Figure 5.** Quantitative analysis of the effects of color flow gain, 20–80, and flow rate, 2–8 mL/s, on velocity estimation and color region area based on TRI data. Solid-line curves demonstrate that increasing flow rate from 2 to 8 mL/s results in a corresponding rise in estimated velocity from approximately 0.5 to 2.0 mm/s. In contrast, variations in color flow gain do not significantly affect the measured velocity. Dotted-line plots represent the number of color pixels within the ROI, which increases with higher flow rates and gradually approaches saturation between 6 and 8 mL/s.

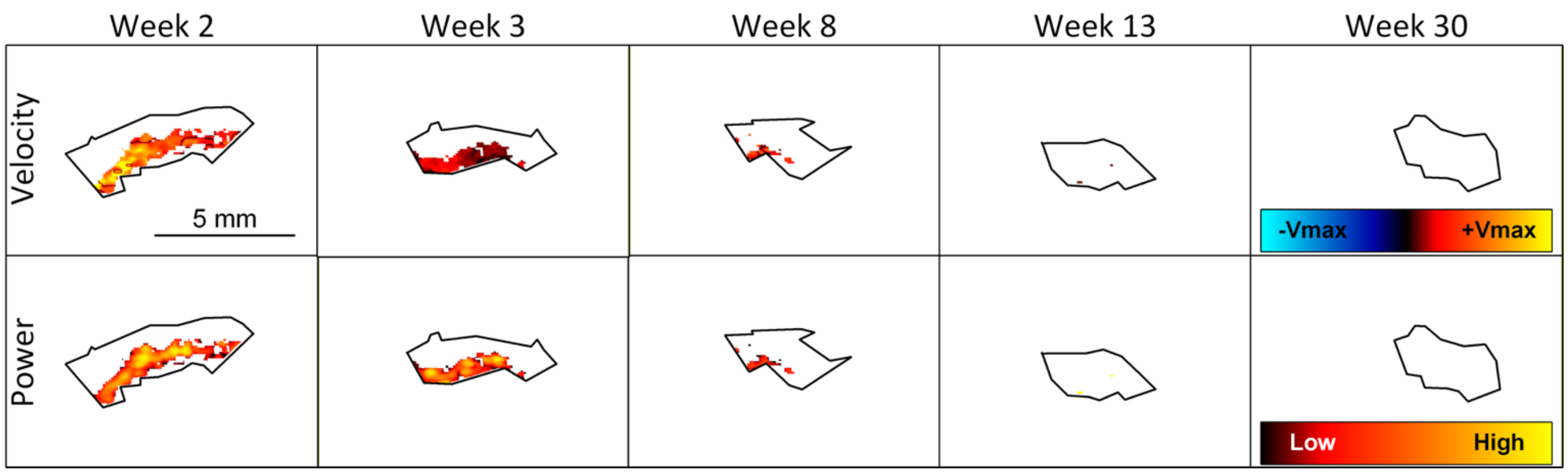

**Figure 6.** Representative results from the developed technique in one patient across multiple visits, showing color-based velocity (top) and power (bottom) color maps. The extent of the color region progressively decreases over time in both the color velocity and power maps. This reduction is hypothesized to be partially due to decreased vascular demand as tissue recovery progresses, resulting in reduced blood flow at later stages. Visit weeks may vary between patients due to scheduling availability.

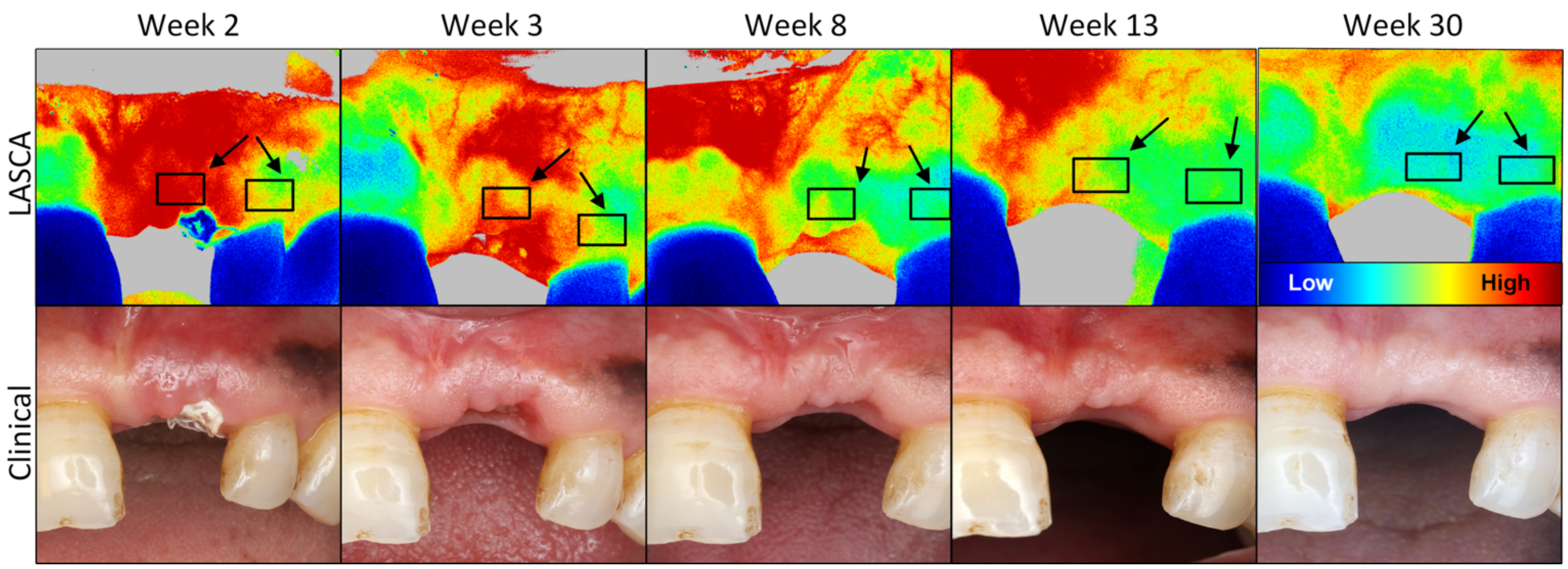

**Figure 7.** Representative LASCA images and corresponding clinical photographs from the same patient illustrate longitudinal changes in tissue perfusion and visual tissue appearance, respectively. The LASCA images show a transition from higher to lower perfusion during the healing period from week 2 to week 30, reflected by a progressive color shift from red to blue. At week 2, the implanted membrane remains visible, and the surrounding granulation tissue appears red, indicating elevated perfusion. By week 30, the membrane is fully overgrown with light pink gingival tissue, consistent with the reduced perfusion observed in the LASCA images. Note that only the regions of interest (ROIs) indicated by black rectangles were included in the analysis, with perfusion quantified as the ratio between the extraction site and adjacent healthy tissue.

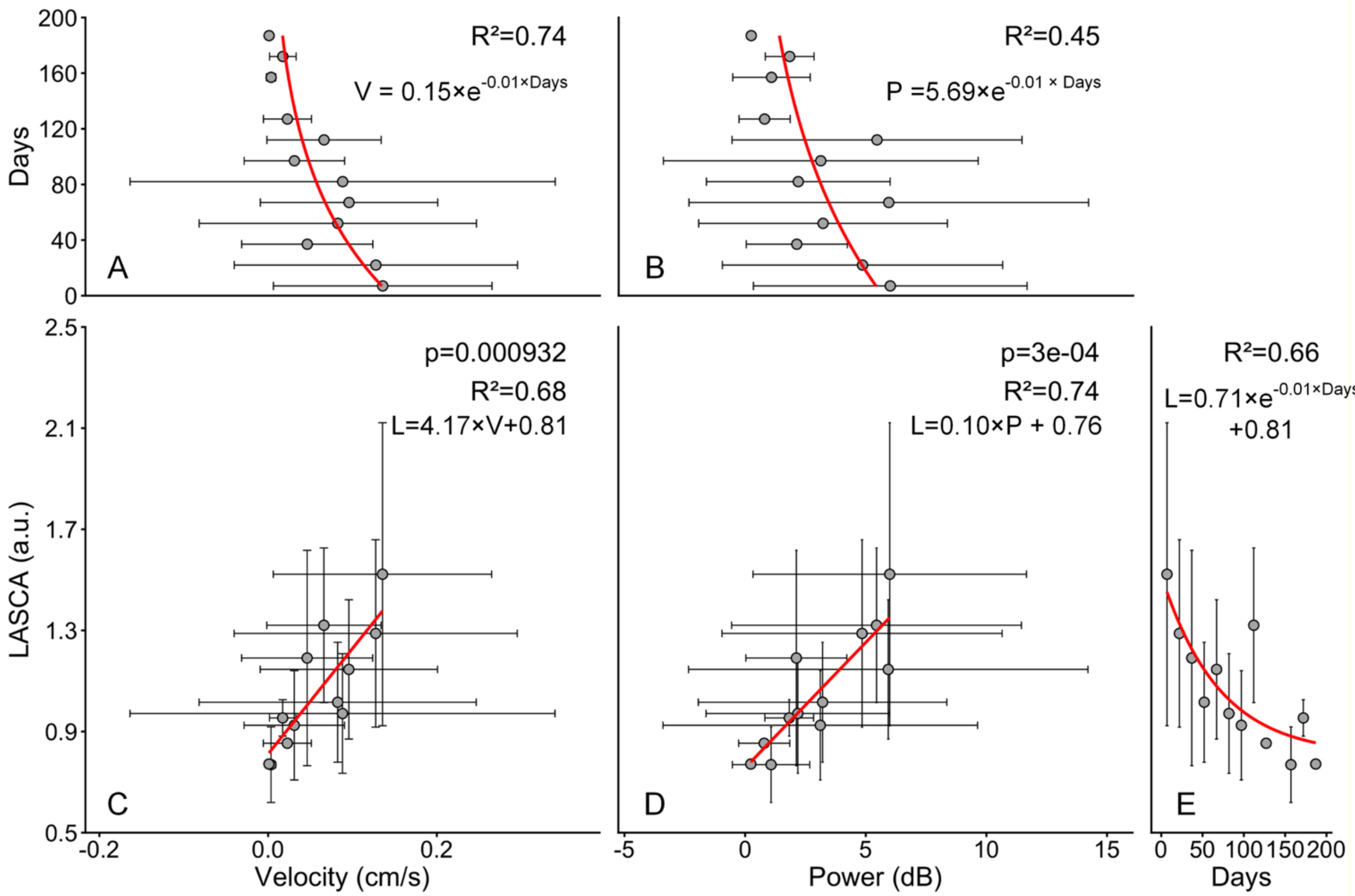

**Figure 8.** Quantitative summary of longitudinal changes in color flow metrics. (A) Nonlinear regression analysis demonstrated a strong temporal association for $CPD_{vel}$ ($R^2$ = 0.74). (B) $CPD_{pwr}$ showed a moderate nonlinear association with time ($R^2$ = 0.45). (C) Linear regression analysis revealed a significant correlation between LASCA and $CPD_{vel}$ ($R^2$ = 0.68, $p < 0.05$). (D) A significant association was also observed between LASCA and $CPD_{pwr}$ ($R^2$ = 0.74, $p < 0.05$). (E) LASCA measurements exhibited a nonlinear temporal trend ($R^2$ = 0.66). Note that panels within the same row share the same y-axis scale, and panels within the same column share the same x-axis scale.